\documentclass[openacc]{rstransa}

\titlehead{Review} 

\jname{rsta}
\Journal{Phil. Trans. R. Soc. A}

\newcommand{\apj}{ApJ}
\newcommand{\aj}{AJ}
\newcommand{\mnras}{MNRAS}

\newcommand{\aap}{A\&A}
\newcommand{\apjs}{ApJS}
\newcommand{\aapr}{AApR}

\newcommand{\pasp}{PASP}

\newcommand{\ssr}{SSR}

\newcommand{\arcsec}{{}^{\prime\prime}}

\newcommand{\refedit}[1]{{{#1}}}

\begin{document}

\makeatother

\title{Strong gravitational lenses
from the Vera C.~Rubin Observatory}

\author{Anowar~J.~Shajib\textsuperscript{1,2,3}, Graham P. Smith\textsuperscript{4}, Simon Birrer\textsuperscript{5}, Aprajita Verma\textsuperscript{6}, Nikki Arendse\textsuperscript{7}, Thomas E.~Collett\textsuperscript{8}, Tansu Daylan\textsuperscript{9}, Stephen Serjeant\textsuperscript{10}, and the LSST Strong Lensing Science Collaboration}

\address{\scriptsize$^1$Department of Astronomy and Astrophysics, 
    University of Chicago,
    Chicago, IL 60637, USA
    
$^2$Kavli Institute for Cosmological Physics,
    University of Chicago,
    Chicago, IL 60637, USA

$^3$NHFP Einstein Fellow

$^4$School of Physics and Astronomy, University of Birmingham, Birmingham, B15 2TT, UK

$^5$Department of Physics and Astronomy, Stony Brook University, Stony Brook, NY 11794, USA
 
$^6$Sub-department of Astrophysics, University of Oxford, Keble Road, Oxford, OX1 3RH, UK

$^7$Oskar Klein Centre, Department of Physics, Stockholm University, SE-106 91 Stockholm, Sweden

$^8$Institute of Cosmology and Gravitation, University of Portsmouth, Burnaby Road, Portsmouth, PO1 3FX, UK

$^9$Department of Physics and McDonnell Center for the Space Sciences, Washington University, St. Louis, MO 63130, USA

$^{10}$School of Physical Sciences, The Open University, Walton Hall, Milton Keynes, MK7 6AA, UK

}

\subject{astrophysics, cosmology, observational astronomy}
\corres{Anowar~J.~Shajib\\ \email{ajshajib@uchicago.edu}}

\begin{abstract}
\footnotesize
Like many areas of astrophysics and cosmology, the Vera C.~Rubin Observatory will be transformational for almost all the applications of strong lensing, thanks to the dramatic increase in the number of known strong lenses by two orders of magnitude or more and the readily available time-domain data for the lenses with transient sources. In this article, we provide an overview of the forecasted number of discovered lenses of different types and describe the primary science cases these large lens samples will enable. We provide an updated forecast on the joint constraint for the dark energy equation-of-state parameters, $w_0$ and $w_a$, from combining all strong lensing probes of dark energy. We update the previous forecast from the Rubin Observatory Dark Energy Science Collaboration's Science Review Document by adding two new crucial strong lensing samples: lensed Type Ia supernovae and single-deflector lenses with measured stellar kinematics. Finally, we describe the current and near-future activities and collaborative efforts within the strong lensing community in preparation for the arrival of the first real dataset from Rubin in early 2026.
\end{abstract}

\begin{fmtext}
\end{fmtext}
\maketitle

\enlargethispage{24pt}

\section{Introduction} \label{sec:intro}

The Vera C.~Rubin Observatory's Legacy Survey of Space and Time (LSST) is a ground-based, time-domain, imaging survey in the optical, which will cover 18,000 square degrees and will operate for ten years \cite{LSSTScienceCollaboration09}. This survey will deliver unprecedented data and impact for many areas in astrophysics and cosmology \cite{Ivezic19}. The field of strong lensing is also set to experience a revolutionary advancement given the increase of known strong lenses to increase by two orders of magnitude or more, thanks to the Rubin LSST \refedit{and other concurrent large-area sky surveys such as \textit{Euclid} and the \textit{Roman} Space Telescope} \cite{Oguri10, Collett15, Ferrami24}. Furthermore, the lenses with time-variable sources will have readily available light curves from the LSST to enable several strong-lensing science applications without requiring follow-up monitoring. Strong lensing systems have many applications in astrophysics and cosmology, each of which will benefit from such a dramatic increase in known systems.

This article overviews some of the main strong lensing science cases to be achieved with the Rubin data. This article is intended to summarise the science cases rather than a detailed review of the methodologies used to accomplish each science goal from the data. However, we refer the readers to comprehensive review articles for the science cases described, with the references provided within relevant sections.

This article is organised as follows. In Section \ref{sec:forecast}, we provide a brief review of the estimates for discoveries of different types of lens systems -- e.g., lensed quasars, lensed supernovae (SNe), galaxy--galaxy lenses. Then, in Section \ref{sec:science_cases}, we describe the main science goals in the field of strong lensing to be accomplished with the Rubin samples. We then describe the current and near-future efforts within the strong-lensing community in preparation for the real Rubin dataset (Data Preview 2) to arrive in 2026 in Section \ref{sec:ongoing_efforts}.

\section{Forecasts for lens discoveries} \label{sec:forecast}

\begin{figure*}[!h]
	\centering
	\includegraphics[width=0.75\textwidth]{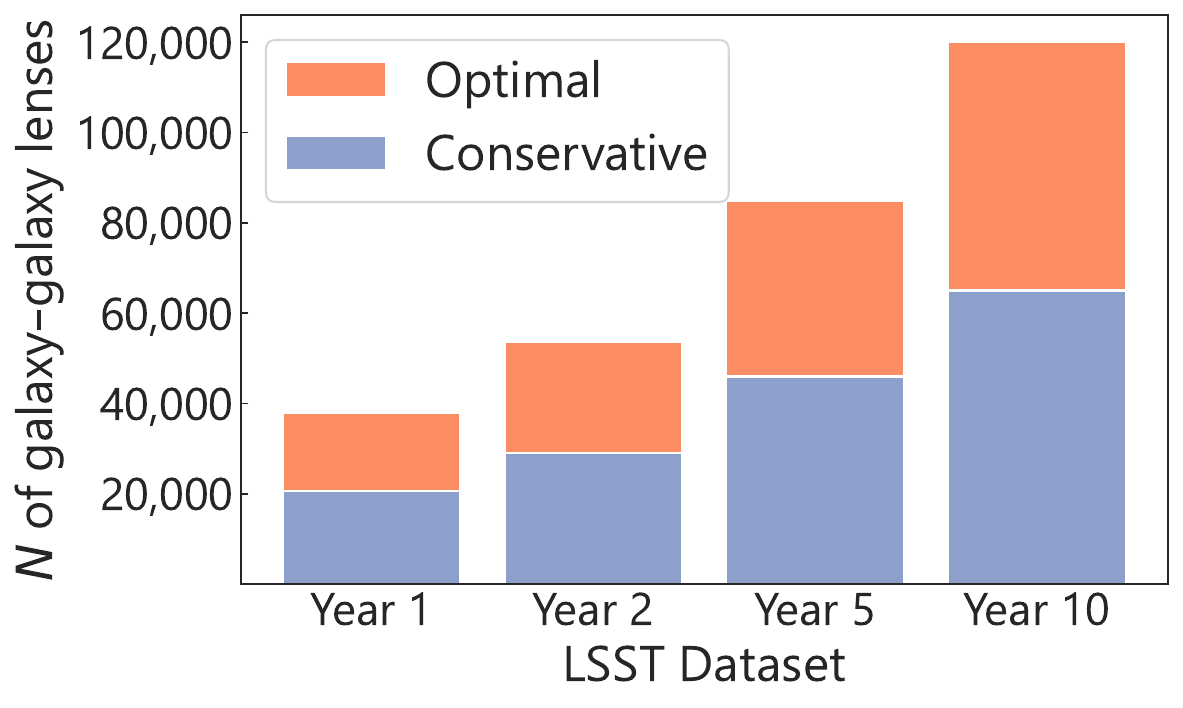}
	\caption{\label{fig:rubin_gg_lens}
	Forecasted number of galaxy--galaxy lenses to be discovered by the Rubin LSST \cite{Collett15}. The numbers for the Year-10 dataset are the directly forecasted ones, and those for the intermediate years are scaled with the imaging signal-to-noise ratio, simply assuming a uniform coverage and cadence, given that the number of discovered lenses linearly scales with the imaging signal-to-noise ratio \cite{Collett15}. The heights of the orange bars show the forecasted numbers for the scenario with optimal image stacking {\cite[for definition]{Collett15}}. The heights of the blue bars represent the `conservative' case where lens finding requires blue-red difference imaging. Year 1--2 datasets will already provide a large fraction (30--45\%) of the total discoverable lenses with the full 10-year dataset.
	}
\end{figure*}

The Rubin Observatory LSST will catalogue 20 billion galaxies after its 10-year survey.\footnote{\url{https://rubinobservatory.org/explore/numbers}} After only one year of the survey, the total size of the catalogue will surpass that of all previous surveys combined. Such a drastic increase in the number of objects will also be reflected in the size of known strong lensing systems. Indeed, \cite{Collett15} predicts 120,000 galaxy-scale lenses to be discovered from the LSST 10-year survey with optimal image stacking, providing two orders of magnitude increase in the number of known galaxy--galaxy lenses (Fig. \ref{fig:rubin_gg_lens}). For the conservative case of lens finding that requires blue-red difference imaging, this number comes down to 62,000 galaxy-scale lenses, still a field-changing number. However, as the LSST will be seeing-limited, the expected strong lens sample will have an Einstein radius distribution that peaks around $1\arcsec$, missing very small separation lenses. While it will be a huge challenge to identify $\mathcal{O}(10^5)$ lenses among $\mathcal{O}(10^{10})$ Rubin galaxies, machine-learning-based search algorithms {\cite{Jacobs19b, Rojas23, Lemon24}} and citizen science \cite{Marshall16} combined with ensemble classifiers {\cite{Holloway24}} \refedit{may} be up to the challenge.

Lensed SNe with galaxy-scale deflectors can be an alternative avenue to detect small separation lenses if discovered through magnification selection. For example, the only two galaxy-scale lensed SNe, iPTF16geu and SN Zwicky, are both small-separation lenses ($\theta_{\rm E} \sim 0.17\arcsec$--$0.3\arcsec$) \cite{Goobar17, Goobar23}. From its 10-year survey, LSST is expected to discover $\sim$380 such unresolved lensed SNe \cite{Goldstein19}, $\sim$180 of which would be Type Ia \cite{SainzdeMurieta23}. In a separate forecast, \cite{Wojtak19} considers discovery methods for both resolved and unresolved lensed SNe and predicts 3400 lensed SNe, a much higher number, to be discovered by the LSST. \cite{Arendse23} has performed a detailed study on the effect of different cadence strategies (e.g., rolling\footnote{For a description, see \url{https://pstn-055.lsst.io/}.} versus non-rolling) on the number of discovered lensed SNIa to find that this number, 44 per year, is largely unaffected by this choice.

For lensed quasars at the galaxy scale, the mock catalogue of \cite{Oguri10} forecasts 3132 systems\footnote{All the forecasted numbers in this paragraph correspond to a 20,000 square degree LSST coverage as assumed by \cite{Yue22} and \cite{Taak23}.} to be discovered after ten years, with a quad fraction of 15\%. From this sample, \cite{Taak23} forecasts that $\sim$1000 lensed quasars will have variability above the LSST photometric uncertainty, making the time delay measurable for use in time-delay cosmography from the LSST light curves alone. \cite{Taak23} additionally demonstrates that a smaller sample of $\sim$250 lenses will have much larger variability for the time delays to be measurable with higher cadence monitoring with smaller 2--4-m class telescopes. \cite{Yue22} provides a new mock catalogue of lensed quasars using recently obtained empirical quasar luminosity functions and the \textsc{CosmoDC2} mock catalogue for the deflector population, forecasting 2377 lensed quasars with 193 quads among them. If this mock catalogue of \cite{Yue22} is used instead of the one from \cite{Oguri10}, the forecasted numbers of systems with measurable time delays from \cite{Taak23} could be lower by a factor of $\sim$5.

The above forecasts were carried out mostly for galaxy-scale lenses, given their relatively simpler population statistics than group- and cluster-scale deflectors. However, qualitatively, the Rubin LSST can be expected to deliver a similar increase by one order of magnitude or more in the number of known group- and cluster-scale lenses \cite{Katsuya24}.

\section{Strong-lensing science cases with Rubin} \label{sec:science_cases}

In this section, we summarise some of the main science goals for strong lensing with the Rubin data or Rubin-discovered lens samples: \ref{sec:cosmo_params} measuring cosmological parameters, \ref{sec:dm_physics} studying dark matter microphysics, \ref{sec:galaxy_evo} studying galaxy evolution at both intermediate ($z \lesssim 1$) and high redshifts ($z \gtrsim 2$), and \ref{sec:transient_physics} studying the structure and physics of transient events. However, this is not a comprehensive list of all the science questions that strong lensing can help answer. The unprecedented increase in the number of known strong lenses delivered by the Rubin LSST will undoubtedly help all the strong lensing science applications.

\subsection{Measuring cosmological parameters} \label{sec:cosmo_params}

Strong lenses can act as a cosmological probe in several ways. First, the time delay between the multiple lensed images carries cosmological information \cite{Refsdal64}. This probe is primarily sensitive to the Hubble constant but also weakly depends on other cosmological parameters that govern late-time expansion history {\cite[for some recent reviews]{Birrer22b, Treu23}}. The time delay can be measured when the background source is either variable or transient. So far, this technique has been applied on eight lensed quasars at the galaxy scale by the Time-Delay COSMOgraphy (TDCOSMO) collaboration \cite{Millon20, Birrer20}, and on three lensed quasars (with time delays measured for six image pairs) on the galaxy-cluster scale \cite{Napier23}. Strongly lensed SNe are much rarer than lensed quasars, with the first such system only being discovered in the 2010s \cite{Quimby13, Kelly15, Goobar17}. Only two lensed SNe have been used for cosmographic measurements, both at the cluster scale \cite{Kelly23, Pascale24}. Although the \textit{JWST} has been impactful in discovering a handful of lensed SNe in recent years {\cite{Pierel24}}, they are serendipitous, and this rate of discovery is not guaranteed to continue in the future. The LSST provides a stable channel for discovering lenses SNe at all deflector scales and will provide a more easily characterisable selection function. We note that to fully harness the cosmological information of the time-delay lenses from the Rubin LSST, follow-up data with other facilities, such as high-cadence monitoring for more precise time delays, high-resolution imaging, and spatially resolved stellar kinematics, will be necessary \cite{Shajib18, Shajib23, Huber19, Ding21}. 

Second, double-source-plane lenses provide a way to constrain a ratio $\beta$ of angular diameter distances ($D$) given by
\begin{equation}
    \beta \equiv \frac{D_{\rm ds1} D_{\rm s2}}{D_{\rm s1} D_{\rm ds2}},
\end{equation}
where d, s1, and s2 in the subscripts of $D$ denote the deflector, source 1, and source 2, respectively \cite{Schneider92}.
Although this ratio is insensitive to the Hubble constant, it is sensitive to the density parameters and the dark energy equation-of-state parameters \cite{Golse02, Jullo10, Collett12}. Whilst many cluster-scale lenses have been discovered with $\ge2$ source planes {\cite[for a review]{Kneib11}}, only a handful of double-source-plane galaxy-scale lenses are known so far \cite{Gavazzi08, Tanaka16, Shajib20}. However, Rubin LSST is expected to deliver a sample of $\mathcal{O}(10^2)$ double-source-plane galaxy-scale lenses \cite{LSST18}. The advantage of galaxy-scale lenses is that their mass distributions are more straightforward to model than group-scale or cluster-scale lenses.

Third, all galaxy-scale lenses, including single-plane ones, with measured kinematics constrain the distance ratio $D_{\rm s}/ D_{\rm ds}$, where $D_{\rm s}$ is the angular diameter distance to the source and $D_{\rm ds}$ is the angular diameter distance between the deflector and the source. Thus, constraining this ratio also allows for measuring the density parameters and the dark energy equation-of-state parameters. \cite{Li24} has forecasted competitive precisions on the cosmological parameters from a sample of 10,000 galaxy-scale lenses (hereafter, single-plane lenses) to be discovered by future surveys such as the LSST, for which single-aperture stellar kinematics would be measured with the 4MOST Strong Lensing Spectroscopic Legacy Survey (4SLSLS)  \cite{Collett23}.

\begin{figure*}
	\includegraphics[width=0.5\textwidth]{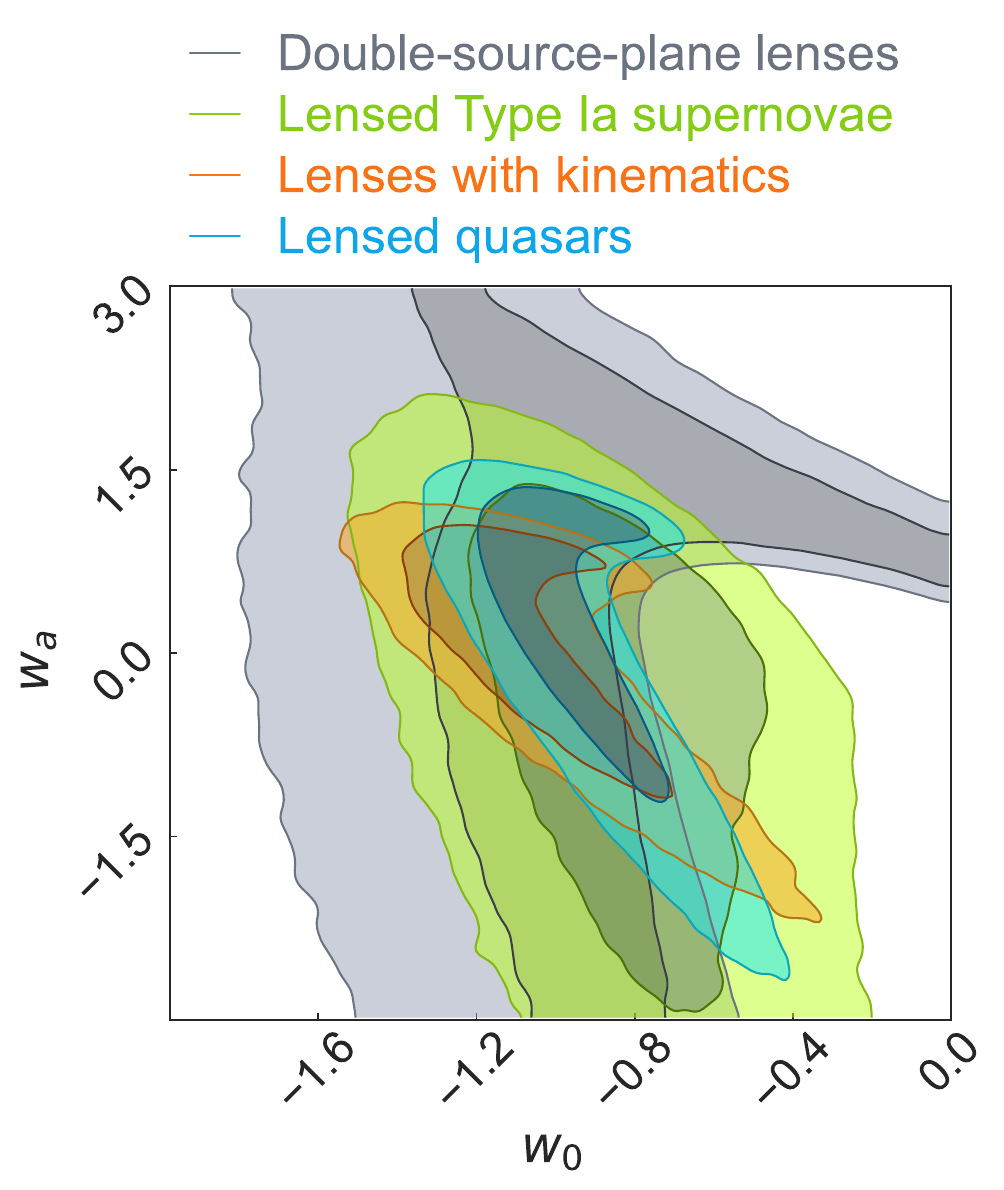}
	\includegraphics[width=0.5\textwidth]{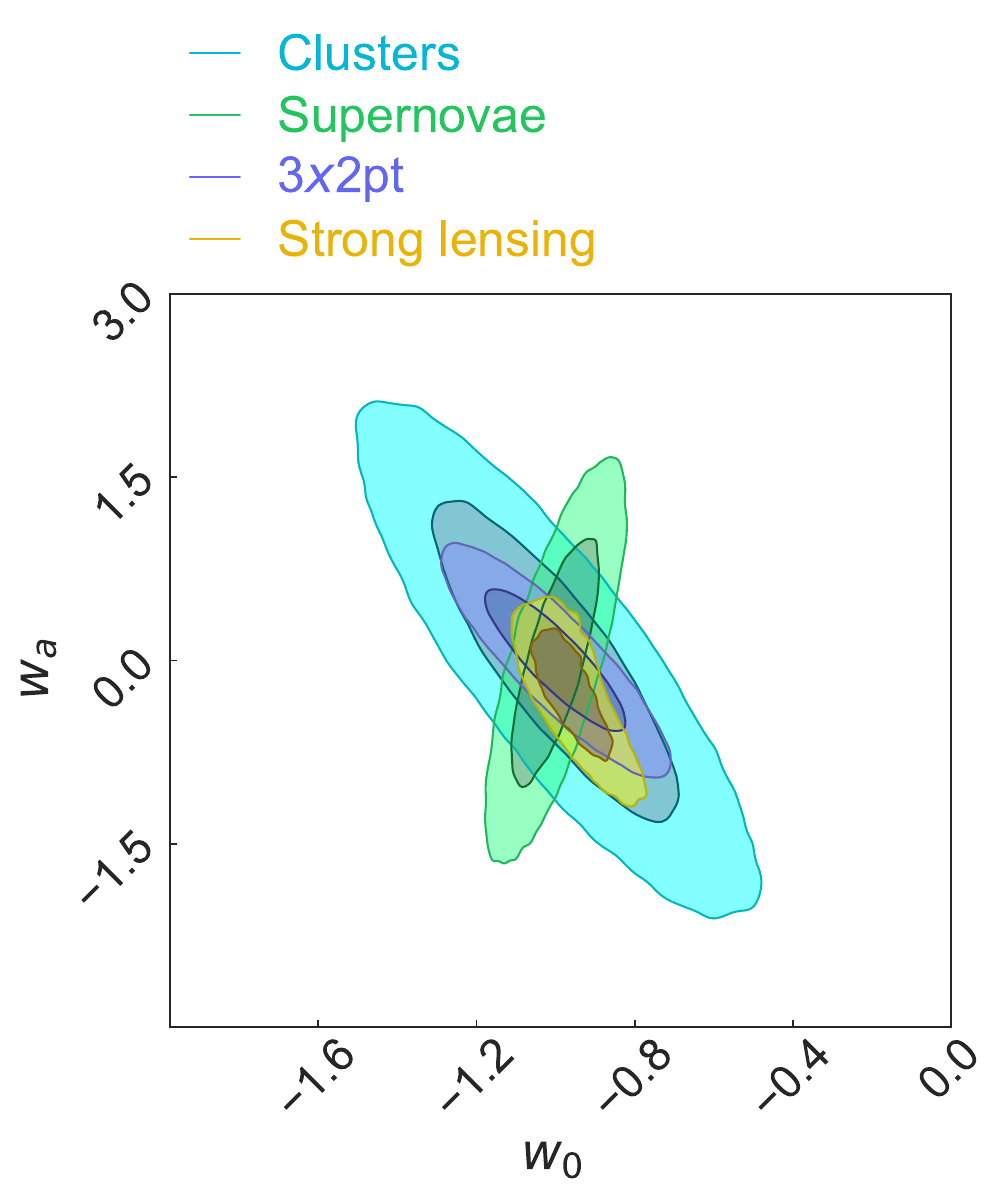}
	\caption{\label{fig:de_forecast}
	The forecasted constraints on the $w_0$--$w_a$ plane from LSST 10-year dataset. \textbf{Left-hand panel:} Individual constraints from the four strong lensing probes. \refedit{The particular shape of degeneracy for a given strong lensing probe originates from using the particular ratio(s) of angular diameter distances to constrain the cosmological parameters.} \textbf{Right-hand panel:} Forecasted joint constraints from combining the four strong lensing probes (yellow), in comparison with three other cosmological probes of Rubin: clusters (cyan), supernovae (green), and 3$\times$2pt correlations (blue). The figures of merit \cite{Albrecht06} for these probes are 11, 32, and 48, respectively. The figure of merit for the combined strong lensing probes is 69, making it the most constraining probe of the dark energy from the Rubin LSST.
	}
\end{figure*}

In this article, we forecast the joint constraint from the above dark energy probes on the equation-of-state parameters, $w_0$ and $w_a$, to be delivered by Rubin LSST samples and follow-up data collected with other facilities. In this forecast, we make a few updates on the previous forecast provided by \cite{LSST18}. First, we include time-delay cosmography with lensed SNIa, for which the mass-sheet degeneracy can be broken with measured magnifications \cite{Birrer22}. Second, we include the sample of single-plane lenses with kinematics from \cite{Li24}. Third, we adopt the number of lensed quasar systems with measured time delays to 236 based on \cite{Taak23} and assume 40 of these systems will have spatially resolved kinematics measured with the \textit{JWST} integral field spectroscopy \cite{Birrer21}. We allow the mass-sheet degeneracy to be broken with the kinematic data for the galaxy--quasar lenses, with full covariance accounted for within the time-delay cosmography sample consisting of lensed quasars and SNIae. We use the same sample of 87 lenses from the mock sample of double-source planes from \cite{LSST18, Sharma23}. We provide additional specifications and settings of the forecast in Appendix \ref{sec:forecast_settings}. We show the forecast for individual probes and compare the joint-strong-lensing forecast with other cosmological probes of the LSST in Fig. \ref{fig:de_forecast}. The combined strong lensing probe will provide the strongest constraint on the dark energy parameters among the Rubin LSST probes, including 3$\times$2 point correlations, SNe, and clusters, based on their current forecasts \cite{LSST18}. Combining strong lensing constraints with those from the other independent probes will constrain the dark energy parameters even tighter due to the orthogonality of the contours.

A caveat of this forecast is that it does not account for correlated systematics between the time-delay cosmography sample and the single-plane lenses. Although accounting for correlated modelling systematics can potentially increase the forecasted uncertainty, the large sample of single-plane lenses with measured kinematics, in turn, can effectively constrain the mass-sheet degeneracy more tightly \cite{Birrer20}. This degeneracy can be broken even more effectively if stacked galaxy--galaxy weak lensing measurements for this large sample of single-plane lenses are also used \cite{Khadka24}. Thus, this sample of single-plane lenses may either increase the constraining power of the time-delay cosmography sample or alleviate the need for relatively expensive spatially resolved stellar kinematics measurement. A detailed exploration of the trade-off between these two counteracting effects is left for a future study.

\subsection{Dark matter microphysics} \label{sec:dm_physics}

\refedit{Strong lensing can be used to explore how dark matter is structured at small scales, which yields a strong probe of the particle nature of dark matter. The abundance, spatial distribution, and $N$-point correlation of dark matter substructure in the halo of lensing galaxies differ for different theories of dark matter, such as cold and warm dark matter \cite[for a review]{Vegetti24}. These substructures can be detected by the `flux-ratio anomaly' in the multiple images \cite{Mao98} or inferring subhalos in the mass model that explains the kinks or perturbations in the lensed arcs or arclets, that is, the `gravitational imaging' method \cite{Koopmans05, Vegetti09, Daylan18, Despali22}. Whereas the gravitational imaging technique can lead to degenerate detection between a substructure within the lens's halo and a line-of-sight halo \cite{Sengul22}, the line-of-sight halos can increase the sensitivity of probing the subhalo mass function with the flux-ratio anomalies \cite{Gilman19}. An alternative method to constrain dark matter microphysics is the power spectrum of dark matter fluctuations \cite{Hezaveh16b}. Unlike the millilensing signature accessible to near-diffraction-limited images of the \textit{Hubble} Space Telescope ({HST}) \cite{Bolton08}, \textit{Roman} \cite{Daylan23}, and the JWST \cite{Keeley24}, the seeing-limited images in LSST will only allow inference of relatively more massive ($M_{\rm sub}\gtrsim10^{11}$ M$_\odot$) substructure in galaxy clusters. In addition, the statistics of substructures within cluster-scale lenses' halos can also be used to test predictions of the cold dark matter paradigm \cite[for a review]{Natarajan24}. The current limitation of these methods is the number of confirmed strong lenses with high signal-to-noise ratios amenable to detailed characterization that can reveal population-level properties across cosmic time while controlling for system-to-system variance. The Rubin LSST is expected to increase the sample size by an order of magnitude.} 

Another path to explore the microphysics of dark matter is through the shape of dark matter halos \cite{Elbert18, Kong24}. The `core--cusp' problem still poses a challenge for the cold dark matter paradigm \cite{deBlok10, Bullock17}, for which galaxy-scale strong lenses can provide key insights, especially for the massive elliptical deflector galaxies that will be detected in abundance by the LSST. In addition, Rubin LSST is expected to discover lenses with dwarf galaxy deflectors, which can be used to probe their halo shape \cite{Drlica-Wagner19}, providing valuable constraints in a relatively understudied mass regime. Additionally, microlensing can be used to constrain the dark matter fraction made of compact objects, such as primordial black holes \cite{Diego22, Arendse24b}.

\subsection{Galaxy evolution} \label{sec:galaxy_evo}

Strong lensing systems allow studying galaxy evolution in two avenues. First, Strong lensing is one of only a handful of probes of the mass distribution in (deflector) galaxies. Thus, we can study the structural evolution of the deflectors shaped by baryonic feedback processes and mergers \cite[for a review]{Shajib22b}. Additionally, comparing the stellar mass measured with lensing--dynamics analysis to that obtained using stellar population synthesis also allows the stellar initial mass function and its evolution to be constrained. \refedit{Whereas some results from this method in the literature have conflicted \cite{Shajib22b}, the unprecedentedly large lens sample from the LSST holds promise to resolve the conflicts, if they originated from small sample size, unaccounted for or non-uniform selection function, or not accounting for correlation with other galaxy properties due to a combination of both of the former reasons.}

Second, strong lensing acts as a `natural telescope'  magnifying distant background galaxies \cite[for a review at the cluster scale]{Natarajan24}. Thus, the structure and spatial properties of these galaxies can be studied at higher spatial resolution compared to unlensed galaxies at the same redshift. Current 10m-class and future extremely large telescopes will play a critical role in such studies. These magnified high-redshift galaxies can, therefore, be used not only for uncovering the impact of baryon cycling in their evolution but also provide an important avenue to answer key questions about cosmic reionisation at the highest redshifts \cite{Treu22}. The large samples of lenses discovered by the Rubin LSST will constrain galaxy properties and their evolution to high precision, both for the deflector and source populations.

\subsection{Structure and physics of transients} \label{sec:transient_physics}

The structure and physics of lensed transient sources can be studied through microlensing, that is, additional magnification or demagnification of a lensed image due to lensing by stars in the foreground lensing galaxy. Microlensing-induced signatures, in the form of `anomalous' flux ratios between multiple images relative to that expected in the absence of microlensing, long-term variability of the flux ratio, or departure from well-modelled SNe light curve shapes, can be identified in the light curves of the images. These signatures allow the structure of the transient source to be probed, for example, quasar accretion disk or SN photosphere \cite[for reviews]{Vernardos24, Suyu24b}. In addition, the statistics of microlensing events probe the stellar IMF in the lensing galaxies \cite{Schechter14}. Thanks to its time-domain nature, Rubin LSST will readily provide light curves of these lensed transients, enabling the long-term variability induced by microlensing to be constrained. These light curves will also contain $\sim$300 per year high-magnification events with much stronger constraining power \cite{Neira20}. Additionally, these light curves can help identify the onset of high-magnification events for higher-cadence monitoring with independent follow-up facilities, with such closely monitored events having been only a handful so far \cite{Vernardos24}.

In addition to the transients mentioned above, strongly lensed gamma-ray bursts, fast radio bursts, kilonovae, and gravitational wave events would be immensely helpful in uncovering the underlying astrophysical mechanisms in these transient events \cite{Oguri19}. For detecting and confirming the lensing nature of these transients, cross-referencing with a catalogue of known lensing objects is a highly promising avenue \cite{Ryczanowski20, Ryczanowski23}. The Rubin LSST will effectively provide the largest such catalogue of lensing systems for this purpose.

\section{Ongoing and near-future community efforts} \label{sec:ongoing_efforts}

The strong lensing community activities are carried out unitedly by two collaborations: the Strong Lensing Science Collaboration\footnote{\url{https://sites.google.com/view/lsst-stronglensing}} (SLSC) and the Dark Energy Science Collaboration\footnote{\url{https://lsstdesc.org/}} (DESC) through its Strong Lensing Topical Team (SLTT)\footnote{SLTT is part of the Time-Domain Working Group in DESC.}. The coordination and exchange of expertise and resources are mutually agreed upon through an official memorandum of understanding.

In preparation for the first light and soon-to-be-delivered first dataset (Data Preview 2 in early 2026) from the Rubin LSST, SLSC and DESC-SLTT are jointly contributing to commissioning Rubin's active optics sub-system and alert production pipeline\footnote{\url{https://sitcomtn-050.lsst.io/}} \cite{Smith21}, and coordinating community efforts in terms of individual data challenges to discover lenses with both static and transient sources. For discovering lenses with static sources from large survey data, machine learning is currently the most popular method \cite[for a review]{Lemon24}. The prerequisite for this kind of data challenge is a realistic simulation pipeline. The \textsc{SLSim} pipeline\footnote{\url{https://github.com/LSST-strong-lensing/slsim}} is currently being developed with that as one of its main goals \cite{Khadka24b}. An additional goal of this pipeline is to characterise the selection function of the discovered lenses, which will be essential to achieve most of the science goals described in Section \ref{sec:science_cases}. Therefore, the deflector and background source galaxy and transient populations are modelled in this pipeline with high realism based on empirical luminosity functions and scaling relations. This pipeline is also being developed with generalizability; thus, it will be easy to use it for similar purposes for \textit{Euclid} or the \textit{Roman} Space Telescope.

\section{Summary}

In this article, we provided an overview of the strong lensing science cases enabled by the Rubin LSST. First, we briefly reviewed the forecasted number of different types of strong lensing systems (Section \ref{sec:forecast}). We then described some of the principal strong-lensing science cases to be achieved with the LSST-discovered strong lensing samples, either utilizing the LSST data alone or with follow-up data obtained from other facilities (Section \ref{sec:science_cases}). We provided an updated forecast on constraining the dark energy equation of state parameters $w_0$ and $w_a$. In this forecast, we demonstrate that by combining the four strong-lens samples -- lensed quasars with measured time delays, lensed SNIae with measured time delays, single-plane lenses with kinematics, and double-source-plane lenses -- strong lensing will become the most constraining dark energy probe for the Rubin LSST. Finally, we described the current and near-future community efforts for timely preparation of the analysis pipelines in anticipation of the near-future arrival of the first real dataset from the Rubin LSST in early 2026 (Section \ref{sec:ongoing_efforts}).

\dataccess{The article has no additional data. The code used in the forecast is publicly accessible on GitHub at \url{https://github.com/ajshajib/dark_energy_forecast_rubin_strong_lensing}.}
\aiuse{We have not used AI-assisted technologies in creating this article.}
\aucontribute{A.J.S.: conceptualization, methodology, formal analysis, visualisation, writing---original draft; G.P.S.: conceptualization, writing---review and editing; S.B.: conceptualization, methodology, writing---review and editing; A.V.: conceptualization, writing---review and editing; N.A.: conceptualization, writing---review and editing; T.E.C.: methodology, writing---review and editing, T.D.: writing--review and editing, S.S: writing--review and editing.
\\
All authors gave ﬁnal approval for publication and agreed to be held accountable for the work performed therein.}
\conflict{We declare we have no competing interests.}
\funding{Support for this work was provided by NASA through the NASA Hubble Fellowship grant HST-HF2-51492 awarded to A.J.S. by the Space Telescope Science Institute, which is operated by the Association of Universities for Research in Astronomy, Inc., for NASA, under contract NAS5-26555. G.P.S. acknowledges support from the Leverhulme Trust and the Science and Technology Facilities Council. T.D. acknowledges support from the McDonnell Center for the Space Sciences at Washington University in St. Louis. T.E.C.~is funded by a Royal Society University Research Fellowship, and this project has received funding from the European Research Council (ERC)
under the European Union’s Horizon 2020 research and innovation
programme (LensEra: grant agreement No 945536).}
\ack{\refedit{This article was internally reviewed and approved by the LSST Strong Lensing Science Collaboration. We also thank the anonymous referee for helpful comments that improved this manuscript.} The authors thank the Royal Society for their support. The authors thank Yoon Chon Taak and Tian Li for providing the lens catalogues and MCMC chains from \cite{Taak23, Li24} that were used in the forecast made in this paper. A.J.S. thanks the `Multi-messenger Gravitational Lensing' scientific meeting organisers for their invitation to present this review at the Royal Society.
\\
This article made use of \textsc{hierArc} (\url{https://github.com/sibirrer/hierarc}), \textsc{emcee} \cite{Foreman-Mackey13}, \textsc{lenstronomy} \cite{Birrer18, Birrer21b}, \textsc{numpy} \cite{Oliphant15}, \textsc{astropy} \cite{AstropyCollaboration13, AstropyCollaboration18, AstropyCollaboration22}, \textsc{Scikit-learn} \cite{Pedregosa11}, \textsc{matplotlib} \cite{Hunter07}, \textsc{ChainConsumer} \cite{Hinton16}, and \textsc{seaborn} \cite{Waskom14}.
}

\bibliographystyle{RS} 

\appendix

\section{Settings for dark energy forecasts} \label{sec:forecast_settings}

Here, we describe the settings used in our forecast for Rubin LSST's strong lensing constraints on the dark energy equation of state parameters. We use \textsc{hierArc}\footnote{\url{https://github.com/sibirrer/hierarc}} to sample from the cosmological parameter posterior with \textsc{emcee} \cite{Goodman10, Foreman-Mackey13}. Below, we provide the specific sample and settings for each of the four strong lensing probes of the dark energy. \refedit{We assume space-based (from the {HST}, {JWST}, \textit{Euclid}, or \textit{Roman}) or adaptive-optics-assisted ground-based imaging \cite{Chen19} to be used for lens modelling for the probes below. In addition to needing spectroscopic redshift measurements for all these probes, measured time delays will be necessary for the probes in \ref{sec:tdc_snia} and \ref{sec:tdc_qso}, with velocity dispersion measurements being crucial for the probes in \ref{sec:tdc_qso} and \ref{sec:spl_with_kin}.}

\subsection{Time-delay cosmography with lensed SNIa} \label{sec:tdc_snia}

We adopt the mock sample of 144 lensed SNIa from \cite{Birrer22}: 120 doubles and 24 quads with eight each in the cusp, cross, and fold configurations. For the lensed SNIa sample, we only use the future \textit{Roman} sample from \cite{Birrer22} to provide a prior constraint on the mean apparent magnitude $\overline{m}_{\rm p}$ at the pivot redshift $z_{\rm pivot} = 0.1$. In this forecast, we do not use the relative expansion history from this future \textit{Roman} sample to extract the constraints only from time-delay cosmography with the lensed SNIa sample.

\subsection{Time-delay cosmography with lensed quasars} \label{sec:tdc_qso}

We adopt a sample of 236 lensed quasars from the catalogue of \cite{Oguri10} with variability amplitude more than three times the uncertainty achieved by LSST photometry, based on the estimation by \cite{Taak23}. This large threshold on the variability amplitude is conservative. Thus, this sample size can still be achieved with a lower threshold if the more conservative numbers based on the catalogue of \cite{Yue22} are considered. For example, 346 lenses from the \cite{Yue22} catalogue will have variability amplitude larger than the LSST photometric uncertainty [Y.~C.~Taak, private communication].

We assume 40 of these 236 systems to have \textit{JWST} kinematics and the rest to have single-aperture kinematics measurements. The deflector and source redshifts of these lenses are obtained from the catalogue of \cite{Oguri10}. We make the same assumptions as \cite{Birrer21} for all the other settings in the forecast, including the resolution and uncertainty specifications of the kinematics. We add a population scatter with $\sigma = 0.025$ in the external convergence ($\kappa_{\rm ext}$) for the lensed quasar systems on top of the settings of \cite{Birrer21}, to be self-consistent with the settings applied on the lensed SNIa sample from \cite{Birrer22}. However, for both the lensed SNIa and quasar sample, the population mean and scatter of external convergence are fixed to their true values, since the magnification or the kinematics breaks the total (i.e., internal and external combined) mass-sheet degeneracy, thus leaving a perfect degeneracy between the mean internal mass-sheet-transform parameter $\overline{\lambda}_{\rm int}$ and the mean external convergence $\overline{\kappa}_{\rm ext}$ if both are freely varied.  Following \cite{Birrer21}, we still assume the 2\% uncertainty for each individual system's measurement of the external convergence that contributes toward the overall uncertainty on $\overline{\lambda}_{\rm int}$.

\subsection{Single-plane lenses with kinematics} \label{sec:spl_with_kin}

We use the posterior for this sample of 10,000 lenses directly from \cite{Li24}. When combining this probe with the other three, we use the kernel density estimate of this posterior as an additional prior term jointly applied on $\Omega_{\rm m}$, $w_0$, and $w_a$.

\subsection{Double-source-plane lenses} \label{sec:dspl}

 We use the same sample of 87 double-source-plane lenses from \cite{LSST18, Sharma23} with the same redshift and $\beta$ values. Although the full LSST dataset is forecasted to contain $\sim$500 double-source-plane lenses, the number 87 corresponds to the ones identified from single-epoch images with the best seeing.

\end{document}